\documentclass[aps,twocolumn,superscriptaddress,showpacs]{revtex4}

\usepackage{graphicx}

\begin{document}

\title{On the nature of spinor Bose-Einstein condensates in rubidium}
\author{N. N. Klausen}
\author{J. L. Bohn}
\author{Chris H. Greene}
\affiliation{Department of Physics and JILA, University of Colorado,
Boulder, CO 80309-0440, USA}
 
\date{\today}
 
\begin{abstract}
We perform detailed close-coupling calculations for the rubidium
isotopes $^{85}$Rb and $^{87}$Rb to ascertain the nature of their
spinor Bose-Einstein condensates. These calculations predict that the
spinor condensate for the spin-1 boson $^{87}$Rb has a ferromagnetic
nature. The spinor condensates for the spin-2 bosons $^{85}$Rb and
$^{87}$Rb, however, are both predicted to be polar. The nature of a
spin-1 condensate hinges critically on the sign of the difference
between the \textit{s}-wave scattering lengths for total spin 0 and 2
while the nature of a spin-2 condensate depends on the values of the
differences between \textit{s}-wave scattering lengths for the total
spin 0, 2 and 4. These scattering lengths were extracted previously
and found to have overlapping uncertainties for all three cases, thus
leaving the nature of the spinor condensates ambiguous. The present
study exploits a refined uncertainty analysis of the scattering
lengths based on recently improved result from experimental work by
Roberts \textit{et al.}, which permits us to extract an unambiguous
result for the nature of the ground state spinor condensates.
\end{abstract}


\maketitle

In a conventional magnetic trap for ultra-cold alkali atoms the spin
degrees of freedom are ``frozen out'' since the atom must be in a
weak-field seeking Zeeman state to be trapped. In an optical trap,
however, the spins of the alkali atoms are essentially free, and all
magnetic substates $|f,m\rangle$ for a given spin $f$ can be
populated. Since the atom-atom interaction depends on spin, these
magnetic substates can be changed in a scattering event. Accordingly,
it is of interest to see how the spins are organized in the ground
state and to explore the nature of the spin-mixing dynamics in an
optically trapped Bose-Einstein condensate.

Multi-component condensates have been formed in magnetic traps. For
instance Ref. \cite{Myatt:97} used a double magneto-optical trap and a
magnetic trap to create condensates in either the $|f=2,m=2\rangle$ or
the $|f=1,m=-1\rangle$ spin state of $^{87}$Rb, and in a mixture of
both by cooling $|1,-1\rangle$ evaporatively and $|2,2\rangle$ via
thermal contact with the $|1,-1\rangle$ atoms. In this case the spin
projections are approximately frozen out because the spin-flip cross
sections in $^{87}$Rb are anomalously small
\cite{Burke:97,Julienne:97,Kokkelmans:97}. By contrast,
Ref. \cite{Stamper-Kurn:98} made a sodium condensate consisting
simultaneously of all three magnetic substates of the $f=1$ atomic
state, by cooling the atoms in a magnetic trap and then transferring
them into a optical trap. This experimental technique produces what is
referred to as a spinor condensate, because it can explore its full
range of spin degrees of freedom. In the theoretical description of
Ref. \cite{Ho:98,Ho:00}, the spinor condensates are classified
according to the relative values of certain characteristic scattering
lengths. Note that alternative theoretical treatments
\cite{Law:98,Pu:00} differ in their detailed predictions concerning the
nature of the spinor BEC ground state. Nevertheless, in this paper we
determine the interaction parameters for spinor condensates of
$^{85}$Rb and $^{87}$Rb which based on
Ref. \cite{Ho:98,Ho:00,Ciobanu:00} fall into the two following
categories:

\textit{Spin-1 atoms} ($^{87}$Rb) Let $F$ be the total spin of two
bosonic spin $f=1$ atoms, and let $a_F$ be the \textit{s}-wave
scattering length for the total spin $F$ symmetry. Since $f=1$, only
$F=0, 2$ are allowed by Bose symmetry for an \textit{s}-wave
collision. The nature of the spin-1 BEC ground state depends
critically on the relative values of $a_0$ and $a_2$. According to Ho
\cite{Ho:98} a spinor Bose condensate composed of spin-1 bosons in an
optical trap can be either \textit{``ferromagnetic''} or
\textit{``antiferromagnetic''} in nature \cite{Ho:98,Ho:00}. The
antiferromagnetic state has alternatively been termed
\textit{``polar''}, and we use this terminology here. The difference
between the scattering lengths $a_0$ and $a_2$ determines the nature
of the spin-1 condensate: the ferromagnetic state emerges when
$a_0>a_2$, whereas the polar state emerges when $a_0<a_2$
\cite{Ho:98}. In the ferromagnetic state virtually all atoms reside in
the same spin substate (either $m=1$ or $m=-1$); in the polar state
the spin projections are mixed.

\textit{Spin-2 atoms} ($^{85}$Rb, $^{87}$Rb) Two bosonic spin $f=2$
atoms possess $F=0$, $2$, $4$ total spin states exhibiting the
appropriate Bose symmetry for an \textit{s}-wave collision. For spin-2
$^{87}$Rb the scattering lengths $a_0$, $a_2$ and $a_4$ are determined
by the real part of the phase-shift since the inelastic scattering
processes are also allowed. According to Ciobanu \textit{et al.}
\cite{Ciobanu:00}, a spinor condensate of spin-2 bosons in an optical
trap can be one of the three types \textit{``ferromagnetic''},
\textit{``polar''}, or \textit{``cyclic''} in nature, which we
abbreviate as F, P or C respectively. Ferromagnetic and polar
condensates are similar to those above. The name ``cyclic'' arises
from a close analogy with \textit{d}-wave BCS superfluids. The nature
of the spin-2 BEC ground state depends critically on the relative
values of $a_0-a_2$ and $a_2-a_4$
\cite{Ciobanu:00}. 

\newpage
The three states emerge under following conditions:

\hspace{.2in}P: $a_0-a_4 < 0$, $\frac{2}{7}\mid a_2-a_4 \mid <
\frac{1}{5} \mid a_0-a_4 \mid$,

\hspace{.2in}F: $a_2-a_4 > 0$, $\frac{1}{5}\mid a_0-a_4 \mid +
\frac{2}{7} (a_2-a_4) > 0$,

\hspace{.2in}C: $a_2-a_4 < 0$, $\frac{1}{5}\mid a_0-a_4 \mid -
\frac{2}{7} (a_2-a_4) > 0$.

For spin-1 $^{87}$Rb the total spin $F=0$ and $F=2$ scattering lengths
$a_0$ and $a_2$ are almost equal. They have been calculated before
\cite{Ho:98} based on the analysis of Ref. \cite{Roberts:98}, but the
uncertainties determined still overlap for $a_0$ and $a_2$, so that
the sign of the difference has remained uncertain. In particular, the
scattering lengths have been interpreted rather conservatively in
Ref. \cite{Roberts:98}. For spin-2 $^{85}$Rb and $^{87}$Rb the
uncertainties for the total spin $F$ scattering lengths $a_0$, $a_2$
and $a_4$ have been too large to uniquely identify the nature of the
spinor condensates
\cite{Ho:00}. The uncertainty region for $^{85}$Rb was large enough to
overlap all three regions P, F and C, while the uncertainty region for
$^{87}$Rb overlapped both the polar and the cyclic region. In the
present study we determine the scattering lengths $a_0$ and $a_2$, and
their uncertainties for spin-1 $^{87}$Rb, and $a_0$, $a_2$ and $a_4$
and their uncertainties for spin-2 $^{85}$Rb and $^{87}$Rb. We
concentrate on an accurate determination of the difference $a_0 - a_2$
for spin-1 $^{87}$Rb and the pair ($a_0-a_4$, $a_2-a_4$) for spin-2
$^{85}$Rb and $^{87}$Rb. If one accepts the spinor condensate
treatment of Ref. \cite{Law:98,Pu:00} this analysis gives an
unambiguous determination of the nature of the BEC ground states.

Uncertainties in the scattering lengths arise primarily from imperfect
knowledge of three parameters: the long-range van der Waals
coefficient $C_6$, and the singlet and the triplet \textit{s}-wave
scattering lengths, $a_s$ and $a_t$ respectively. In addition, when
using potential curves determined for one isotope to predict
scattering for another isotope, the results can depend on the precise
number of bound states in the triplet potential, $N_b$ as well as the
precise number of bound states in the singlet potential. Roberts
\textit{et al.}  analyzed a magnetic-field Feshbach resonance to
determine ``state of the art'' potentials for $^{85}$Rb
\cite{Roberts:98}. Recently they have revisited some of the
rethermalization measurements in Ref. \cite{Roberts:98} and improved
the uncertainties for the long-range van der Waals coefficient and the
singlet end triplet
\textit{s}-wave scattering lengths for $^{85}$Rb \cite{Roberts:01}. 

Using these new values of $C_6$, $a_s$ and $a_t$ we show below
unambiguously that $a_0 > a_2$ for spin-1 $^{87}$Rb. This result in
turn implies that the spinor condensate is definitely ferromagnetic,
as was previously suspected \cite{Ho:98}. By contrast the spin-1
$^{23}$Na scattering lengths, recently determined in
Ref. \cite{Crubellier:99}, imply that a $^{23}$Na $f=1$ spinor BEC is
polar, as has been suggested before \cite{Ho:98}. By extracting the
scattering length from a spectroscopic experiment, Crubellier
\textit{et al.} found that, for $^{23}$Na, $a_0=50.0\pm1.6$ $a.u.$ and
$a_2=55.0\pm1.7$ $a.u.$ \cite{Crubellier:99}. They calculated the
scattering lengths for two values of the $C_6$ coefficient for
$^{23}$Na and found that the influence of the $C_6$ value is very
small (a 4\% change in $C_6$ results in a variation in the scattering
length of the order of 0.1\%). Consequently, the analysis for
$^{23}$Na \cite{Crubellier:99}, in conjunction with the present
analysis for $^{87}$Rb, implies that both types of spin-1 condensates
can be realized with the atoms used most frequently in BEC experiments
($^{23}$Na and $^{87}$Rb).

\begin{figure}
\includegraphics[width=\columnwidth]{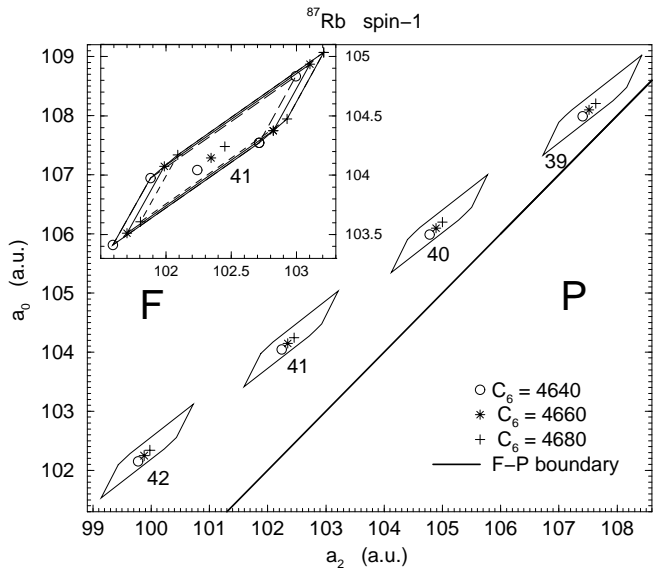}
\caption{Total spin $F=0$ scattering length versus total spin $F=2$
scattering length for spin-1 $^{87}$Rb. The uncertainties of $a_0$ and
$a_2$ are determined by the uncertainties on $a_s$, $a_t$, $C_6$ and
$N_b$, the number of bound states in the $^{85}$Rb triplet
potential. The symbols in the middle of the ``diamonds'' are the mean
scattering length for each $C_6$ and the diamonds encircle the
uncertainties arising from uncertainties on $a_s$ and $a_t$ for all
$C_6$. The thick black line shows the boundary between the
ferromagnetic and polar phase of the spinor condensate: $a_0 = a_2$
and the number next to each diamond is $N_b$.}
\label{fig87_1}
\end{figure}

\begin{figure}
\includegraphics[width=\columnwidth]{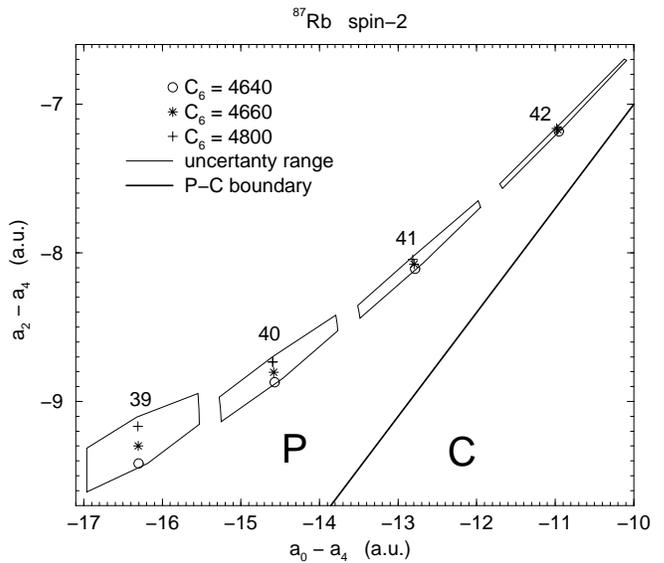}
\caption{Difference between total spin $F=2$ and $F=4$ scattering
lengths versus the difference between total spin $F=0$ and $F=4$
scattering lengths for spin-2 $^{87}$Rb. The uncertainties of $a_0$,
$a_2$ and $a_4$ are determined by the uncertainties of $a_s$, $a_t$,
$C_6$ and $N_b$, the number of bound states in the $^{85}$Rb triplet
potential. The symbols in the middle of the ``diamonds'' are the mean
scattering length for each $C_6$ and the diamonds encircle the
uncertainties arising from uncertainties on $a_s$ and $a_t$ for all
$C_6$. The thick black line shows the boundary between the polar and
the cyclic phase of the spinor condensate: $(a_2-a_4) =
\frac{-7}{10}(a_2-a_4)$ and the number next to each diamond is $N_b$.}
\label{fig87_2}
\end{figure}

\begin{figure}
\includegraphics[width=\columnwidth]{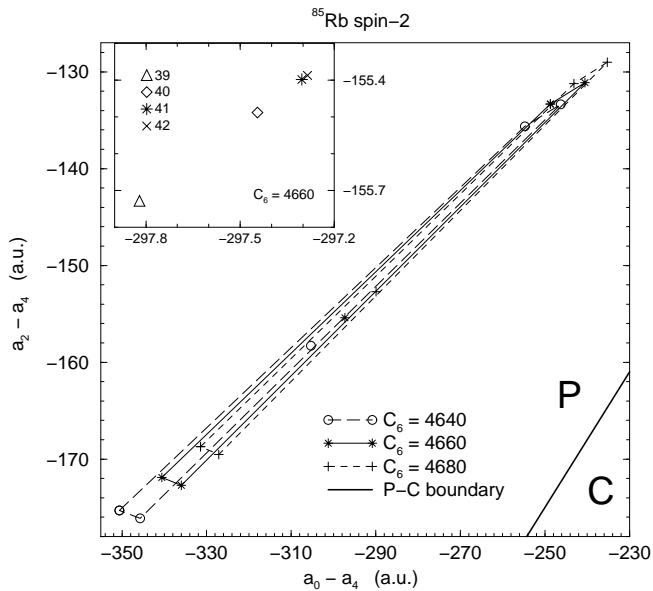}
\caption{Difference between total spin $F=2$ and $F=4$ scattering
lengths versus the difference between total spin $F=0$ and $F=4$
scattering lengths for spin-2 $^{85}$Rb. The uncertainties of $a_0$,
$a_2$ and $a_4$ are determined by the uncertainties of $a_s$, $a_t$,
$C_6$ and $N_b$, the number of bound states in the $^{85}$Rb triplet
potential. The symbols in the middle of the ``diamonds'' are the mean
scattering length for $N_b=41$ and each $C_6$ and the diamonds
encircle the uncertainties arising from uncertainties on $a_s$ and
$a_t$ for each $C_6$. The thick black line shows the boundary between
the polar and the cyclic phase of the spinor condensate: $(a_2 -
a_4)=\frac{-7}{10}(a_2 - a_4)$ and the close-up shows the average
values for the four different $N_b$.}
\label{fig85_2}
\end{figure}

The improved results for $C_6$, $a_s$ and $a_t$ also predict that $a_0
- a_4 < 0$, $\frac{2}{7}\mid a_2-a_4 \mid < \frac{1}{5} \mid a_0-a_4
\mid$, for both spin-2 $^{85}$Rb and $^{87}$Rb. This result implies that
both spin-2 $^{85}$Rb and $^{87}$Rb will be polar. Previously, it was
estimated that $^{87}$Rb would be polar, but that $^{87}$Rb would be
cyclic \cite{Ciobanu:00}. This implies that the ground state for
spin-2 $^{85}$Rb and $^{87}$Rb will have the same nature as spin-2
$^{23}$Na. Spin-2 $^{23}$Na was already unambiguously classified since
the uncertainties on differences between the relevant scattering
lengths place $a_0 - a_4$ and $a_2-a_4$ within the polar region
\cite{Ciobanu:00}. The results for spin-1 $^{87}$Rb and for spin-2
$^{85}$Rb and $^{87}$Rb are summarized in Figs. \ref{fig87_1},
\ref{fig87_2} and \ref{fig85_2} respectively.

Our calculations start from the singlet and triplet Born-Oppenheimer
potentials between two rubidium atoms that were calculated in
Ref. \cite{Krauss:90}, where the sing-let potential is adjusted to have
125 bound states \cite{private}. These potentials are matched smoothly
at $r =22.0$ $a.u.$ to the standard long-range van der Waals
potentials using the new value of the long-range coefficient $C_6$
inferred from the experiment in Ref. \cite{Roberts:98} and reanalyzed
according to Ref. \cite{Roberts:01}, and using the $C_8$ and $C_{10}$
coefficients from the calculations of Ref. \cite{Marinescu:94}. The
potentials are adjusted to match the scattering length by including
short-range inner-wall corrections that are parameterized for each
spin by: $c\arctan((r-r_{min})^2/(c_r))$ for $r<r_{min}$. $c_r$ is a
constant (the same order of magnitude as $r_{min}$; slightly different
for the singlet and the triplet), the inner-wall parameters $c$ are of
the order of $10^{-5}$ to $10^{-4}$ $a.u.$, $r$ is the separation
between the two Rb atoms and $r_{min}$ is the separation for which the
potential is minimal. The inner-wall parameters $c$ are varied over a
range that reproduces the recently improved values of $a_s$ and $a_t$.
The improved values of $C_6$ for rubidium, $a_s$ and $a_t$ for
$^{85}$Rb are: $C_6 = 4660\pm20$ $a.u.$, $a_s = 3650^{+1500}_{-670}$
$a.u.$ and $a_t = -332\pm18$ $a.u.$ \cite{Roberts:01}, while the
calculations of Ref. \cite{Marinescu:94} determined that $\bar{C}_8 =
550600$ $a.u.$. These are the values we adopt in the present
calculations. Our calculations here do not allow for variance in
$C_8$. This is reasonable because the dependence of $C_8$ is one order
of magnitude smaller than the dependence of $C_6$. Furthermore the
number of bound states in the triplet potential was previously
believed to be $39\pm1$ \cite{private,Tsai:97}, but more refined
experimental analysis suggests that it is instead $40 \leq N_b \leq
42$ \cite{private,Leo}. The present calculations are done for
$N_b=39,40,41,42$. The number of bound states in the singlet potential
is not changed.

Our calculations have been carried out for three values of $C_6$ that
span the empirical range ($4640-4680$ $a.u.$). These values are
adequate since the quantities of interest vary smoothly with $C_6$
over the range of interest. We also tested the triplet potential for
each one of the four relevant $N_b$ ($39$, $40$, $41$, $42$). For each
value of $N_b$ we determine the values of the inner-wall corrections
that correspond to the uncertainty range of $^{85}$Rb $a_s$ and $a_t$
for each of the three values of $C_6$. These calculations are carried
out at zero magnetic field and 130 $nK$ since the given value of
$C_6$, $a_s$ and $a_t$ are determined from collisions at this
temperature
\cite{Roberts:01}. The same potentials optimized for $^{85}$Rb have
been used in our $^{87}$Rb calculations, except for an appropriate
change in the reduced mass. Since $N_b$ in rubidium is unknown at
present, and since we utilize the same potentials determined by the
$^{85}$Rb singlet and triplet scattering length in our $^{87}$Rb
calculations, we incorporate $N_b$ in our analysis of the
uncertainties for $^{87}$Rb. The singlet and triplet potentials are
used in multichannel calculations to compute the $a_0$ and $a_2$ for
spin-1 $^{87}$Rb and $a_0$, $a_2$ and $a_4$ for spin-2 $^{85}$Rb and
$^{87}$Rb (again at zero magnetic field and at $130$ $nK$), with fixed
$C_6$ and $N_b$. The calculations are then repeated for each value of
$C_6$ and $N_b$ with the corresponding new values of the inner-wall
corrections. These calculations span the empirical $C_6$, $N_b$, $a_s$
and $a_t$ range, which permits us to extract the over-all uncertainty
in the difference between the relevant scattering lengths, $a_0-a_2$
for spin-1 and the two relevant differences ($a_0-a_4$, $a_2-a_4$) for
spin-2.

As a confirmation, the same rubidium potentials have also been used to
calculate the singlet and triplet \textit{s}-wave scattering lengths
$a_s$ and $a_t$ for $^{87}$Rb (at 130 $nK$ energy). These
single-channel calculations have been repeated for the three values of
$C_6$ and four values of $C_6$ that span the empirical range. This
permits us to check whether $a_s$ and $a_t$ fall within the range of
previous measured values. The single-channel triplet scattering
lengths are found to be:

\( \begin{array}{cc}
	N_b & a_t \\
	39 & 107\pm1\ a.u. \\
	40 & 103\pm1\ a.u. \\
	41 & 100\pm1\ a.u. \\
	42 & 97\pm1\ a.u.
\end{array} \)

\noindent and the single-channel singlet scattering length is found to
be: $a_s = 91\pm1$ $a.u.$. The $a_s$ and $a_t$ values for $N_b=39$
are in good agreement with previous work
\cite{Roberts:98,Burke:99}. As another confirmation, and since $N_b$
is unknown we have also calculated the $^{85}$Rb scattering length for
$f=2$, $m=-2$ at various magnetic fields to compare the obtained
values with the values from Roberts \textit{et al.};
\cite{Roberts:01}. This comparison shows good agreement. For each of
the $N_b$, the scattering lengths obtained for the specific magnetic
fields exhibits an uncertainty greater than the one given by
\cite{Crubellier:99}.

The values for $a_0$ and $a_2$ for spin-1 $^{87}$Rb,along with their
uncertainties, are shown in Fig. \ref{fig87_1}. $a_0$ is always
greater than $a_2$ in the multichannel calculations, which
unambiguously determine the nature of spin-1 $^{87}$Rb to be
ferromagnetic. The global difference lies between $0.3$ and $2.7$
$a.u.$ over the uncertainty range. The difference is an increasing
function of $N_b$, while $a_0$ and $a_2$ themselves are decreasing
functions of $N_b$. For a given $N_b$, the range of possible values of
$a_0$ and $a_2$ varies only weakly with $C_6$, as was the case for
$^{23}$Na \cite{Crubellier:99}.

The results of $a_0-a_4$ and $a_2-a_4$ with uncertainties for
$^{87}$Rb spin-2 are shown in Fig. \ref{fig87_2}. For all four values
of $N_b$ and all three values of $C_6$ the uncertainty region is
within the ``polar'' region, making the nature of $^{87}$Rb spin-2
condensate unambiguously determined. The pair ($a_0-a_4$, $a_2-a_4$)
moves closer to the boundary between the polar and the cyclic regions
as $N_b$ is increased but never reaches the boundary, within the
present uncertainties. For a fixed value of $N_b$ $a_2-a_4$ is
increasing as a function of $C_6$, while $a_0-a_4$ is almost
independent of $C_6$. The uncertainty region for a fixed value of
$N_b$ is very narrow (especially for higher $N_b$). The long axis of
this region corresponds to the difference $a_s - a_t$, whereas the
narrow axis corresponds to the sum $a_s + a_t$.

The results for $^{85}$Rb spin-2 are shown on Fig. \ref{fig85_2}. In
contrast to the case of $^{87}$Rb, here $a_2-a_4$ and $a_0-a_4$ are
more dependent on the value of $C_6$ than on $N_b$, but only very
little. $a_2-a_4$ and $a_0-a_4$ are slowly increasing functions of
$C_6$ as well as of $N_b$. The uncertainties for all $N_b$ and values
of $C_6$ unambiguously determine the nature of $^{85}$Rb spin-2 to be
polar. The uncertainty region is again very narrow. The long axis of
this region corresponds to $a_t$, whereas the narrow axis corresponds
to $a_s$.

Since the graphs for spin-2 $^{85}$Rb and $^{87}$Rb show only
scattering lengths differences rather than scattering lengths, we
summarize $a_0$, $a_2$ and $a_4$ in the following table. The
scattering lengths for $^{85}$Rb show only very little dependence of
$N_b$. Over the entire range of $a_s$, $a_t$, $C_6$ and $N_b$ the
estimated scattering lengths for $^{85}$Rb are (in $a.u.$):

$a_0=-680\pm80$, $a_2=-540\pm50$, $a_4=-390\pm30$

\noindent The scattering lengths for spin-2 $^{87}$Rb over the entire
range of $a_s$, $a_t$ and $C_6$ are estimated as (in $a.u.$):

\( \begin{array}{ccccc}
	N_b & a_0 & a_2 & a_4 \\
39 & 90.5\pm1 & 97.5\pm1 & 106.8\pm1 \\
	40 & 89.0\pm1 & 94.8\pm1 & 103.6\pm1 \\
	41 & 87.7\pm1 & 92.4\pm1 & 100.5\pm1 \\
	42 & 86.4\pm1 & 90.2\pm1 & 97.4\pm1
\end{array} \)

\noindent These numbers conservatively give the global uncertainties
for each $N_b$. In the context of spinor condensates it is necessary
to consider the actual allowed regions of the parameters, as we have
done above and which permit us to draw meaningful conclusions.

To see how the results change when the multichannel and single-channel
energy changes, we calculated the scattering lengths at various
energies (with 1 pico Kelvin in the single and multichannel as the
lowest value) to cover the relevant temperature for some
experiments. This did not change our conclusions about the nature of
the spinor Bose-Einstein condensates in rubidium.

Since the shape of the inner-wall potential is not known exactly and
since we change it to have potentials with the four different values
of $N_b$ we also performed the calculations with a quadratic
inner-wall correction ($c(r-r_{min})^2$ for $r<r_{min}$) instead of
the $\arctan$ form. This did not change the conclusions and only
changed the calculated scattering lengths by about $0.1$ \%.

The present values of $a_0$, $a_2$ for spin-1 $^{87}$Rb and $a_0$,
$a_2$, $a_4$ for spin-2 $^{85}$Rb and $^{87}$Rb are consistent with
values obtained from Ref. \cite{Roberts:98}. Note that to carry out
the calculations based on Ref. \cite{Roberts:98}, the correlations
among $a_s$, $a_t$ and $C_6$ must be taken into account. We have
separately calculated the values of $a_0$, $a_2$ for spin-1 $^{87}$Rb
and $a_0$, $a_2$, $a_4$ for spin-2 $^{85}$Rb and $^{87}$Rb form $a_s$,
$a_t$ for $^{85}$Rb and $C_6$ as given in Ref. \cite{Roberts:98}, and
find that they support our classifications of the spinor condensates
as presented in this paper. The new values from Ref. \cite{Roberts:01}
allow us to determine a smaller uncertainty on the calculated
scattering lengths, but they do not change our conclusions.

In summary our analysis based on the new results for the values of
$C_6$, $a_s$, $a_t$ and the number of bound states in the triplet
potential demonstrate that the nature of the ground states of
$^{85}$Rb and $^{87}$Rb spin-2 condensates should be polar. In
addition, the ground state of the $^{87}$Rb spin-1 condensate should
be ferromagnetic. Therefore, in view of the known scattering
parameters for $^{23}$Na, both ferromagnetic and polar spin-1
condensates are experimental accessible whereas no cyclic or
ferromagnetic spin-2 condensate appears to exist for the most common
rubidium isotopes.

We gratefully acknowledge D. M. Stamper-Kurn for suggesting that we
perform a refined uncertainty analysis for spin-1 $^{87}$Rb. we also
thank E. Snyder for permitting the use of his computer programs,
J. P. Burke for assistance and valuable discussions and J. L. Roberts
for communication the results of Ref. \cite{Roberts:01} prior to
publication. This work is supported by the National Science
Foundation. N.N.K. acknowledges support from the Danish Fulbright
Commission.


\begin{thebibliography}{10}
\expandafter\ifx\csname bibnamefont\endcsname\relax
  \def\bibnamefont#1{#1}\fi
\expandafter\ifx\csname bibfnamefont\endcsname\relax
  \def\bibfnamefont#1{#1}\fi
\expandafter\ifx\csname url\endcsname\relax
  \def\url#1{\texttt{#1}}\fi
\expandafter\ifx\csname urlprefix\endcsname\relax\def\urlprefix{URL }\fi
\providecommand{\bibinfo}[2]{#2}
\providecommand{\eprint}[2][]{\url{#2}}

\bibitem{Myatt:97}
\bibinfo{author}{\bibfnamefont{C.~J.} \bibnamefont{Myatt}},
  \bibinfo{author}{\bibfnamefont{E.~A.} \bibnamefont{Burt}},
  \bibinfo{author}{\bibfnamefont{R.~W.} \bibnamefont{Ghrist}},
  \bibinfo{author}{\bibfnamefont{E.}~\bibnamefont{Cornell}}, \bibnamefont{and}
  \bibinfo{author}{\bibfnamefont{C.~E.} \bibnamefont{Wieman}},
  \bibinfo{journal}{Phys. Rev. Lett.}
  \textbf{\bibinfo{volume}{78}}(\bibinfo{number}{4}), \bibinfo{pages}{586}
  (\bibinfo{year}{1997}).

\bibitem{Burke:97}
\bibinfo{author}{\bibfnamefont{J.~P.} \bibnamefont{Burke}},
  \bibinfo{author}{\bibfnamefont{J.~L.} \bibnamefont{Bohn}},
  \bibinfo{author}{\bibfnamefont{B.~D.} \bibnamefont{Esry}}, \bibnamefont{and}
  \bibinfo{author}{\bibfnamefont{C.~H.} \bibnamefont{Greene}},
  \bibinfo{journal}{Phys. Rev. A}
  \textbf{\bibinfo{volume}{55}}(\bibinfo{number}{4}), \bibinfo{pages}{R2511}
  (\bibinfo{year}{1997}).

\bibitem{Julienne:97}
\bibinfo{author}{\bibfnamefont{P.~S.} \bibnamefont{Julienne}},
  \bibinfo{author}{\bibfnamefont{F.~H.} \bibnamefont{Mies}},
  \bibinfo{author}{\bibfnamefont{E.}~\bibnamefont{Tiesinga}}, \bibnamefont{and}
  \bibinfo{author}{\bibfnamefont{C.~J.} \bibnamefont{Williams}},
  \bibinfo{journal}{Phys. Rev. Lett.}
  \textbf{\bibinfo{volume}{78}}(\bibinfo{number}{10}), \bibinfo{pages}{1880}
  (\bibinfo{year}{1997}).

\bibitem{Kokkelmans:97}
\bibinfo{author}{\bibfnamefont{S.~J. J. M.~F.} \bibnamefont{Kokkelmans}},
  \bibinfo{author}{\bibfnamefont{H.~M. J.~M.} \bibnamefont{Boesten}},
  \bibnamefont{and} \bibinfo{author}{\bibfnamefont{B.~J.}
  \bibnamefont{Verhaar}}, \bibinfo{journal}{Phys. Rev. A}
  \textbf{\bibinfo{volume}{55}}(\bibinfo{number}{3}), \bibinfo{pages}{R1589}
  (\bibinfo{year}{1997}).

\bibitem{Stamper-Kurn:98}
\bibinfo{author}{\bibfnamefont{D.~M.} \bibnamefont{Stamper-Kurn}},
  \bibinfo{author}{\bibfnamefont{M.~R.} \bibnamefont{Andrews}},
  \bibinfo{author}{\bibfnamefont{A.~P.} \bibnamefont{Chikkatur}},
  \bibinfo{author}{\bibfnamefont{S.}~\bibnamefont{Inouye}},
  \bibinfo{author}{\bibfnamefont{H.-J.} \bibnamefont{Miesner}},
  \bibnamefont{and} \bibinfo{author}{\bibfnamefont{J.}~\bibnamefont{Stenger}},
  \bibinfo{journal}{Phys. Rev. Lett.}
  \textbf{\bibinfo{volume}{80}}(\bibinfo{number}{10}), \bibinfo{pages}{2027}
  (\bibinfo{year}{1998}), \bibinfo{note}{see also \cite{Roberts:98,Esry:97}}.

\bibitem{Ho:98}
\bibinfo{author}{\bibfnamefont{T.-L.} \bibnamefont{Ho}},
  \bibinfo{journal}{Phys. Rev. Lett.}
  \textbf{\bibinfo{volume}{81}}(\bibinfo{number}{4}), \bibinfo{pages}{742}
  (\bibinfo{year}{1998}).

\bibitem{Ho:00}
\bibinfo{author}{\bibfnamefont{T.-L.} \bibnamefont{Ho}} \bibnamefont{and}
  \bibinfo{author}{\bibfnamefont{S.~K.} \bibnamefont{Yip}},
  \bibinfo{journal}{Phys. Rev. Lett.}
  \textbf{\bibinfo{volume}{84}}(\bibinfo{number}{18}), \bibinfo{pages}{4031}
  (\bibinfo{year}{2000}).

\bibitem{Law:98}
\bibinfo{author}{\bibfnamefont{C.~K.} \bibnamefont{Law}},
  \bibinfo{author}{\bibfnamefont{H.}~\bibnamefont{Pu}}, \bibnamefont{and}
  \bibinfo{author}{\bibfnamefont{N.~P.} \bibnamefont{Bigelow}},
  \bibinfo{journal}{Phys. Rev. Lett.}
  \textbf{\bibinfo{volume}{81}}(\bibinfo{number}{24}), \bibinfo{pages}{5257}
  (\bibinfo{year}{1998}).

\bibitem{Pu:00}
\bibinfo{author}{\bibfnamefont{H.}~\bibnamefont{Pu}},
  \bibinfo{author}{\bibfnamefont{C.~K.} \bibnamefont{Law}}, \bibnamefont{and}
  \bibinfo{author}{\bibfnamefont{N.~P.} \bibnamefont{Bigelow}},
  \bibinfo{journal}{Physica B} \textbf{\bibinfo{volume}{280}},
  \bibinfo{pages}{27} (\bibinfo{year}{2000}).

\bibitem{Ciobanu:00}
\bibinfo{author}{\bibfnamefont{C.~V.} \bibnamefont{Ciobanu}},
  \bibinfo{author}{\bibfnamefont{S.-K.} \bibnamefont{Yip}}, \bibnamefont{and}
  \bibinfo{author}{\bibfnamefont{T.-L.} \bibnamefont{Ho}},
  \bibinfo{journal}{Phts. Rev. A}
  \textbf{\bibinfo{volume}{61}}(\bibinfo{number}{3}),
  \bibinfo{pages}{033607(5)} (\bibinfo{year}{2000}).

\bibitem{Roberts:98}
\bibinfo{author}{\bibfnamefont{J.~L.} \bibnamefont{Roberts}},
  \bibinfo{author}{\bibfnamefont{N.~R.} \bibnamefont{Claussen}},
  \bibinfo{author}{\bibfnamefont{J.~P.} \bibnamefont{Burke}},
  \bibinfo{author}{\bibfnamefont{C.~H.} \bibnamefont{Greene}},
  \bibinfo{author}{\bibfnamefont{E.~A.} \bibnamefont{Cornell}},
  \bibnamefont{and} \bibinfo{author}{\bibfnamefont{C.~E.}
  \bibnamefont{Wieman}}, \bibinfo{journal}{Phys. Rev. Lett.}
  \textbf{\bibinfo{volume}{81}}(\bibinfo{number}{23}), \bibinfo{pages}{5109}
  (\bibinfo{year}{1998}).

\bibitem{Roberts:01}
\bibinfo{author}{\bibfnamefont{J.~L.} \bibnamefont{Roberts}},
  \bibinfo{author}{\bibfnamefont{J.~P.} \bibnamefont{Burke}},
  \bibinfo{author}{\bibfnamefont{N.~R.} \bibnamefont{Claussen}},
  \bibinfo{author}{\bibfnamefont{S.~L.} \bibnamefont{Cornish}},
  \bibinfo{author}{\bibfnamefont{E.~A.} \bibnamefont{Donley}},
  \bibnamefont{and} \bibinfo{author}{\bibfnamefont{C.~E.}
  \bibnamefont{Wieman}}, \bibinfo{journal}{Phys. Rev. A}
  (\bibinfo{year}{2001}), \bibinfo{note}{submitted}.

\bibitem{Crubellier:99}
\bibinfo{author}{\bibfnamefont{A.}~\bibnamefont{Crubellier}},
  \bibinfo{author}{\bibfnamefont{O.}~\bibnamefont{Dulieu}},
  \bibinfo{author}{\bibfnamefont{F.}~\bibnamefont{Masnou-Seeuws}},
  \bibinfo{author}{\bibfnamefont{M.}~\bibnamefont{Elbs}},
  \bibinfo{author}{\bibfnamefont{H.}~\bibnamefont{Knockel}}, \bibnamefont{and}
  \bibinfo{author}{\bibfnamefont{E.}~\bibnamefont{Tiemann}},
  \bibinfo{journal}{European Physical Journal D}
  \textbf{\bibinfo{volume}{6}}(\bibinfo{number}{2}), \bibinfo{pages}{211}
  (\bibinfo{year}{1999}).

\bibitem{Krauss:90}
\bibinfo{author}{\bibfnamefont{M.}~\bibnamefont{Krauss}} \bibnamefont{and}
  \bibinfo{author}{\bibfnamefont{W.~J.} \bibnamefont{Stevens}},
  \bibinfo{journal}{J. Chem. Phys.}
  \textbf{\bibinfo{volume}{93}}(\bibinfo{number}{6}), \bibinfo{pages}{4236}
  (\bibinfo{year}{1990}).

\bibitem{private}
\bibinfo{author}{\bibfnamefont{J.~P.} \bibnamefont{Burke}}
  (\bibinfo{year}{2000}), \bibinfo{note}{private correspondence}.

\bibitem{Marinescu:94}
\bibinfo{author}{\bibfnamefont{M.}~\bibnamefont{Marinescu}},
  \bibinfo{author}{\bibfnamefont{H.~R.} \bibnamefont{Sadeghpour}},
  \bibnamefont{and} \bibinfo{author}{\bibfnamefont{A.}~\bibnamefont{Dalgarno}},
  \bibinfo{journal}{Phys. Rev. A} \textbf{\bibinfo{volume}{49}},
  \bibinfo{pages}{982} (\bibinfo{year}{1994}).

\bibitem{Tsai:97}
\bibinfo{author}{\bibfnamefont{C.~C.} \bibnamefont{Tsai}},
  \bibinfo{author}{\bibfnamefont{R.~S.} \bibnamefont{Freeland}},
  \bibinfo{author}{\bibfnamefont{J.~M.} \bibnamefont{Vogels}},
  \bibinfo{author}{\bibfnamefont{H.~M. J.~M.} \bibnamefont{Boesten}},
  \bibinfo{author}{\bibfnamefont{B.~J.} \bibnamefont{Verhaar}},
  \bibnamefont{and} \bibinfo{author}{\bibfnamefont{D.~J.}
  \bibnamefont{Heinzen}}, \bibinfo{journal}{Phys. Rev. Lett.}
  \textbf{\bibinfo{volume}{79}}(\bibinfo{number}{7}), \bibinfo{pages}{1245}
  (\bibinfo{year}{1997}).

\bibitem{Leo}
\bibinfo{author}{\bibfnamefont{P.}~\bibnamefont{Leo}} \bibnamefont{and}
  \bibinfo{author}{\bibfnamefont{E.}~\bibnamefont{Tiesinga}},
  \bibinfo{note}{unpublished}.

\bibitem{Burke:99}
\bibinfo{author}{\bibfnamefont{J.~P.} \bibnamefont{Burke}} \bibnamefont{and}
  \bibinfo{author}{\bibfnamefont{J.~L.} \bibnamefont{Bohn}},
  \bibinfo{journal}{Phys. Rev. A}
  \textbf{\bibinfo{volume}{59}}(\bibinfo{number}{2}), \bibinfo{pages}{1303}
  (\bibinfo{year}{1999}).

\bibitem{Esry:97}
\bibinfo{author}{\bibfnamefont{B.~D.} \bibnamefont{Esry}},
  \bibinfo{author}{\bibfnamefont{C.~H.} \bibnamefont{Greene}},
  \bibinfo{author}{\bibfnamefont{J.~P.} \bibnamefont{Burke}}, \bibnamefont{and}
  \bibinfo{author}{\bibfnamefont{J.~L.} \bibnamefont{Bohn}},
  \bibinfo{journal}{Phys. Rev. Lett.}
  \textbf{\bibinfo{volume}{19}}(\bibinfo{number}{19}), \bibinfo{pages}{3594}
  (\bibinfo{year}{1997}).

\end{thebibliography}

\end{document}